\documentclass[aps,prd,onecolumn,nofootinbib]{revtex4-2}

\usepackage{amsmath,amssymb}
\usepackage{graphicx}
\usepackage[colorlinks=true, linkcolor=blue, citecolor=blue, urlcolor=blue]{hyperref}
\usepackage{physics}
\usepackage{slashed}
\usepackage{lipsum} 
\usepackage{subcaption}
\newcommand{\Trh}{T_\text{rh}}

\begin{document}

\title{Freeze-in at Low Reheating and Direct Detection of Fermion Dark Matter}

\author{Amir Amiri}
\affiliation{Department of Physics, Faculty of Science, Ferdowsi University of Mashhad, P.O.Box 1436,
Mashhad, Iran}

\author{Bastian Diaz Saez}
\affiliation{Millennium Institute for Subatomic Physics at the High-Energy Frontier (SAPHIR)\\
Fernández Concha 700, Santiago, Chile}
\affiliation{Instituto de Física, Pontificia Universidad Católica de Chile\\
Avenida Vicuña Mackenna 4860, Santiago, Chile}

\author{Kilian Möhling}
\affiliation{Institut für Kern- und Teilchenphysik, TU Dresden, Zellescher Weg 19, 01069 Dresden, Germany}

\date{\today}

\begin{abstract}
We investigate a low–reheating-temperature freeze-in scenario within a minimal model of fermionic dark matter interacting through a pseudoscalar mediator. In this setup, dark matter is produced via the decay of the pseudoscalar, which remains in thermal equilibrium with the Standard Model bath. We derive the thermalization and non-thermalization conditions for the new fields and obtain the corresponding direct-detection constraints and projections on the model based on LZ and DARWIN experiments, respectively.
\end{abstract}

\maketitle

\section{Introduction}
Freeze-in \cite{Hall:2009bx} is an attractive mechanism for generating dark matter (DM) abundance in the early Universe. In this scenario, the DM particle interacts only feebly with the Standard Model (SM), such that it never attains thermal equilibrium. Instead, its abundance is gradually accumulated through decays or annihilations of particles in the thermal bath. The production rate depends either on the interaction strength between DM and a particle in equilibrium (typically a mediator) or on the maximum temperature reached by the thermal bath, the \textit{reheating temperature} $\Trh$.

If $\Trh$ is low compared to the mass scales relevant to DM production \cite{Elahi:2014fsa, Frangipane:2021rtf}, the couplings controlling the freeze-in process can be several orders of magnitude larger than in the standard high-$\Trh$ case. This opens up new opportunities to probe such scenarios, most notably through direct detection experiments \cite{Cosme:2023xpa, Silva-Malpartida:2023yks, Boddy:2024vgt, Koivunen:2024vhr, Arcadi:2024wwg, Bernal:2024ndy, Belanger:2024yoj, Mondal:2025awq,  Roy:2025moo, Bernal:2025fcl}.

In this work, we consider a minimal Higgs-portal scenario in which fermionic DM interacts with a pseudoscalar mediator via a Yukawa coupling. Similar setups have been studied in the context of high-$\Trh$ freeze-in \cite{Yaguna:2023kyu, Yin:2024sle}, where the correct relic abundance can be reproduced, but the parameter space remains entirely beyond the reach of direct detection. Here, instead, we focus on the low-reheating regime, in which DM is produced predominantly through the decay of the pseudoscalar mediator. We also analyze the direct detection prospects of this framework and derive the corresponding constraints from current and future experiments.

The structure of the paper is as follows. In Sec.~\ref{sec:model} we introduce the minimal setup. Section~\ref{sec:relic} describes the computation of the relic abundance at low $\Trh$, including freeze-in production from pseudoscalar decays and comments on the possible super-WIMP contribution \cite{Feng:2003xh}. In Sec.~\ref{sec:thermal}, we derive the thermalization and non-thermalization conditions for the pseudoscalar mediator and the fermionic DM. Direct detection constraints are presented in Sec.~\ref{sec:dd}, and we summarize our conclusions in Sec.~\ref{sec:conclusions}.

\section{The model}
\label{sec:model}
We extend the Standard Model (SM) by two gauge-singlet fields: a Dirac fermion $\psi$ and a real scalar $s$. We adopt the following choice: we impose a CP symmetry acting on the new fields as $\psi \rightarrow \gamma^0\psi$ and $s \rightarrow -s$ \cite{DiazSaez:2021pmg, Belyaev:2022qnf}. Under this symmetry, the DM kinetic and interaction Lagrangian is
\begin{equation}
\mathcal{L} \;=\;
\bar{\psi}\left(i\slashed{\partial} - m_\psi\right)\psi
\;+\;
\frac{1}{2}\partial_\mu s\,\partial^\mu s 
\;-\; 
\frac{\mu_s^2}{2} s^2
\;+\;
i y_p\, s\, \bar{\psi}\gamma^5\psi
\;-\;
\lambda_s s^4
\;-\;
\lambda_{hs} (H^\dagger H) s^2
\end{equation}
Here we chose $\mu_s^2>0$, such that only the Higgs doublet acquires a vacuum expectation
value $\ev{H}=(0,\frac{v}{\sqrt2})$ (with $v\approx 246$ GeV). 
After EWSB, the physical pseudoscalar mass given by $m_s^2 = \mu_s^2 + \frac{\lambda_{hs} v^2}{2}$.
The relevant parameters for DM phenomenology are the scalar and fermion singlet masses $m_s$ and $m_\psi$,
the DM Yukawa coupling $y_p$ and the quartic singlet--Higgs coupling $\lambda_{hs}$.
For completeness we have also included the quartic singlet self-coupling $\lambda_s$ which, however, has negligible 
impact on both DM production and direct detection and will therefore be ignored in the following.
The scan ranges for the parameters considered in this work are summarized in Table~\ref{tab:scan_ranges}.
\begin{table}[t]
    \centering
    \begin{tabular}{c c}
    \hline
    Parameter & Scan range \\
    \hline
    $m_\psi$ & $[1{ \rm GeV},\, 1{\rm TeV}]$ \\
    $m_{s}$ & $[2{\rm GeV},\, 1{\rm TeV}]$ \\
    $y_p$ & $[10^{-9},\, \pi]$ \\
    $\lambda_{hs}$ & $1$ (fixed) \\
    \hline
    \end{tabular}
    \caption{Parameter ranges used in our numerical scan. The Higgs portal coupling value $\lambda_{hs}$ was fixed for simplicity, while the remaining parameters were varied within the indicated intervals.}
    \label{tab:scan_ranges}
\end{table}
  In particular, we focus on the regime $m_s > 2m_\psi$ where the fermion $\psi$ is the only cosmologically stable particle and thus plays the role of the DM candidate, while $s$ behaves as an unstable mediator. For $\lambda_{hs} \lesssim \mathcal{O}(\pi)$, the pseudoscalar thermalizes efficiently with the SM bath (as long as $\Trh$ is not too small), while the Yukawa coupling $y_p$ must remain sufficiently small to keep $\psi$ out of equilibrium, ensuring freeze-in production.

If instead $m_s < 2 m_\psi$, the pseudoscalar cannot decay into DM particles and becomes cosmologically stable. In that case, both $s$ and $\psi$ contribute to the relic abundance, resulting in a two-component DM scenario \cite{DiazSaez:2021pmg}, which we do not consider here.

\section{DM production at low $\Trh$}\label{sec:relic}
We focus on the scenario in which the dark matter particle $\psi$ is produced via the freeze-in mechanism \cite{Hall:2009bx} through the decay of the pseudoscalar mediator $s$ in a Universe with a low reheating temperature $\Trh$. We assume that $s$ remains in thermal equilibrium with the SM bath due to its sizable Higgs-portal coupling $\lambda_{hs}$, while $\psi$ interacts only feebly with both $s$ and the SM, ensuring that it never thermalizes.

Strictly speaking, one should follow the number densities of both $s$ and $\psi$ by solving a coupled system of Boltzmann equations. However, because $s$ is assumed to remain in equilibrium throughout the relevant epoch, its abundance can be safely approximated by its equilibrium distribution. This simplifies the analysis to a single Boltzmann equation governing the production of $\psi$.

For completeness, we also comment at the end of this section on the possible contribution of the super-WIMP mechanism \cite{Feng:2003xh}, in which $s$ freezes out and subsequently decays into DM, providing an additional non-thermal source of $\psi$.
\subsection{Freeze-in production}
The Boltzmann equation that governs the evolution of the DM number density $n_\psi$ is given by \cite{Hall:2009bx}
\begin{eqnarray}
 \frac{dn_\psi}{dt} + 3Hn_\psi = \mathcal{C}\, ,
\end{eqnarray}
where the Hubble parameter $H$ is defined as
\begin{eqnarray}
 H(T) = \frac{\pi}{3}\sqrt{\frac{g_\star(T)}{10}}\frac{T^2}{M_\text{P}},
\end{eqnarray}
with $g_\star(T)$ the effective number of relativistic degrees of freedom at temperature $T$, and $M_{\rm P} \approx 2.4 \times 10^{18}\,\text{GeV}$ the reduced Planck mass. The right-hand side (RHS) of the Boltzmann equation contains the integrated collision term $\mathcal{C}$, which accounts for all processes that produce or deplete the DM abundance. In our setup, the dominant contribution arises from the decay of the pseudoscalar mediator $s$:
\begin{eqnarray}
    s \rightarrow \psi + \bar\psi.
\end{eqnarray}
The collision term can be expressed as\footnote{
Here we have neglected the backreaction term $\sim n_{\psi}n_{\bar\psi}$ due to the small fermion number densities
during freeze-in. However, this term will be relevant for deriving the fermion (non-)thermalization conditions
and will be discussed in more detail below.
}
\begin{eqnarray}
 \mathcal{C} \approx  \Gamma_{s \rightarrow \bar{\psi}\psi} \frac{K_1(m_s/T)}{K_2(m_s/T)} n_s^\text{eq}, \nonumber
\end{eqnarray}	
with $\Gamma_{s \rightarrow \bar{\psi}\psi}$ being the decay width of $s$ into $\bar{\psi}\psi$: 
\begin{eqnarray}
    \Gamma_{s \rightarrow \bar{\psi}\psi} = \frac{y_p^2\, m_s}{8\pi} \left(1 - \frac{4m_\psi^2}{m_s^2}\right)^{1/2}\,,
\end{eqnarray}
$K_1$ and $K_2$ the modified Bessel functions of the second kind, and $n_s^\text{eq}$ is the equilibrium number density of the pseudoscalar mediator
\begin{eqnarray}
 n_s^\text{eq}(T) = g_s\frac{m_s^2}{2\pi^2} T K_2\left(\frac{m_s}{T}\right),
\end{eqnarray}
where $g_s$ is the internal degrees of freedom of the pseudoscalar. 

Introducing the comoving number density $Y_\psi(T) = n_\psi(T)/s(T)$, with $s(T) = \frac{2\pi^2}{45}g_{s*}T^3$ the entropy density, the Boltzmann equation can be rewritten as	
\begin{equation} \label{yield}
  Y_\psi(T) \simeq \int_{T_0}^{\Trh} dT'\, \frac{\mathcal{C}(T')}{H(T')\, s(T')\, T'}\,,
\end{equation}
with $T_0$ the present temperature of the universe and $\Trh$ the reheating temperature. Combining the above equations
yields the comoving DM number density today as
\begin{eqnarray}
\begin{split}
     Y_\psi(T_0) &= \frac{45}{4\pi^5}\frac{g_s \Gamma_{s \rightarrow \bar{\psi}\psi}M_\text{P}}{g_{\star s}} \sqrt{\frac{45\pi m_s^3}{ g_\star}}
     \times \int_{T_0}^{\Trh} dx\, x^{-9/2} e^{-m_s/x} \\
      &\approx \frac{45}{4\pi^5}\frac{g_s \Gamma_{s \rightarrow \bar{\psi}\psi}M_\text{P}}{g_{\star s}} \sqrt{\frac{45\pi m_s}{g_\star}} \frac{e^{-m_s/\Trh}}{\Trh^{5/2}}.   
\end{split}
\end{eqnarray}
Here we used $\Trh \ll m_s$ and approximated $g_{(s)*}\approx g_{(s)*}(\Trh)$. We evaluate $g_{*}$ and $g_{*s}$ at $T_{\rm rh}$ because, in the regime $T_{\rm rh}\ll m_s$, the integrand is exponentially dominated by temperatures near $T_{\rm rh}$ through the factor $e^{-m_s/T}$. Contributions from lower temperatures are negligible, while $g_{*}(T)$ and $g_{*s}(T)$ vary only slowly, so treating them as constants over this narrow range is a good approximation.

Thus, with $g_s = 1$, the DM yield is given by 
\begin{eqnarray}
 Y_\psi(T_0) \approx 0.0035 \; M_\text{P} \; y_p^2 m_s^{3/2} \left(1 - \frac{4m_\psi^2}{m_s^2}\right)^{1/2} 
 \frac{ e^{-m_s/\Trh}}{g_{\star s}\sqrt{g_\star}\, \Trh^{5/2}}
\end{eqnarray}
and the relic density of DM can then be calculated using the usual relation
\begin{eqnarray}
 \Omega_\psi = \frac{m_\psi s_0}{\rho_c} Y_\psi(T_0),
\end{eqnarray}
where $s_0$ is the current entropy density, $Y_\psi(T_0)$ is the yield at the present temperature, and $\rho_c$ is the critical density of the universe.

\subsection{Super-WIMP contribution}
There is an additional contribution to the relic abundance from the so-called super-WIMP mechanism \cite{Feng:2003xh}, in which the pseudoscalar $s$ freezes out and subsequently decays into DM particles. The total relic abundance is then given by the sum of the freeze-in and super-WIMP contributions.

Since we assume that $\lambda_{hs}$ is sufficiently large for $s$ to thermalize, its freeze-out abundance can be computed using standard methods. The super-WIMP contribution to the DM relic density is \cite{Garny:2018ali}
\begin{equation}
\Omega_\psi^{\rm SW} h^2 = \frac{m_\psi}{m_s}\, \Omega_s h^2,
\end{equation}
where $\Omega_s h^2$ denotes the relic density of the pseudoscalar prior to its decay. For example, with $\lambda_{hs}=1$, the super-WIMP contribution remains subdominant ($\lesssim10$\% of the total relic density) for $m_s \lesssim 1~\text{TeV}$, further suppressed by the factor $m_\psi/m_s$. However, for higher values of $m_s$, the super-WIMP contribution can become significant and should be included in the total DM abundance. Here we focus on the regime
where this contribution can be neglected and the DM relic density arises predominantly from the freeze-in mechanism.

\section{Thermalization}\label{sec:thermal}
As the thermal evolution of the different species starts at a low reheating temperature, the interaction rates among the particles tend to be suppressed. Therefore, we must ensure that the freeze-in production of $\psi$ via the in-equilibrium $s$ satisfies the basic requirements of the setup: a thermalized pseudoscalar and a non-thermal DM particle. In the Appendix \ref{app_termal}, we present the derivation of the conditions for (non-)thermalization in each case.

\subsection{Pseudoscalar thermalization} \label{subsec:pseudo_therm}
At the reheating temperature $\Trh$, we require that the pseudoscalar mediator $s$ be in thermal equilibrium with the SM particles. The dominant process driving the thermalization of $s$ is the annihilation of SM particles into $s$ pairs via the Higgs-portal coupling $\lambda_{hs}$. The corresponding interaction rate $\Gamma_s$ for this process can be estimated as

\begin{figure}[t]
	\centering
\includegraphics[width=0.4\textwidth]{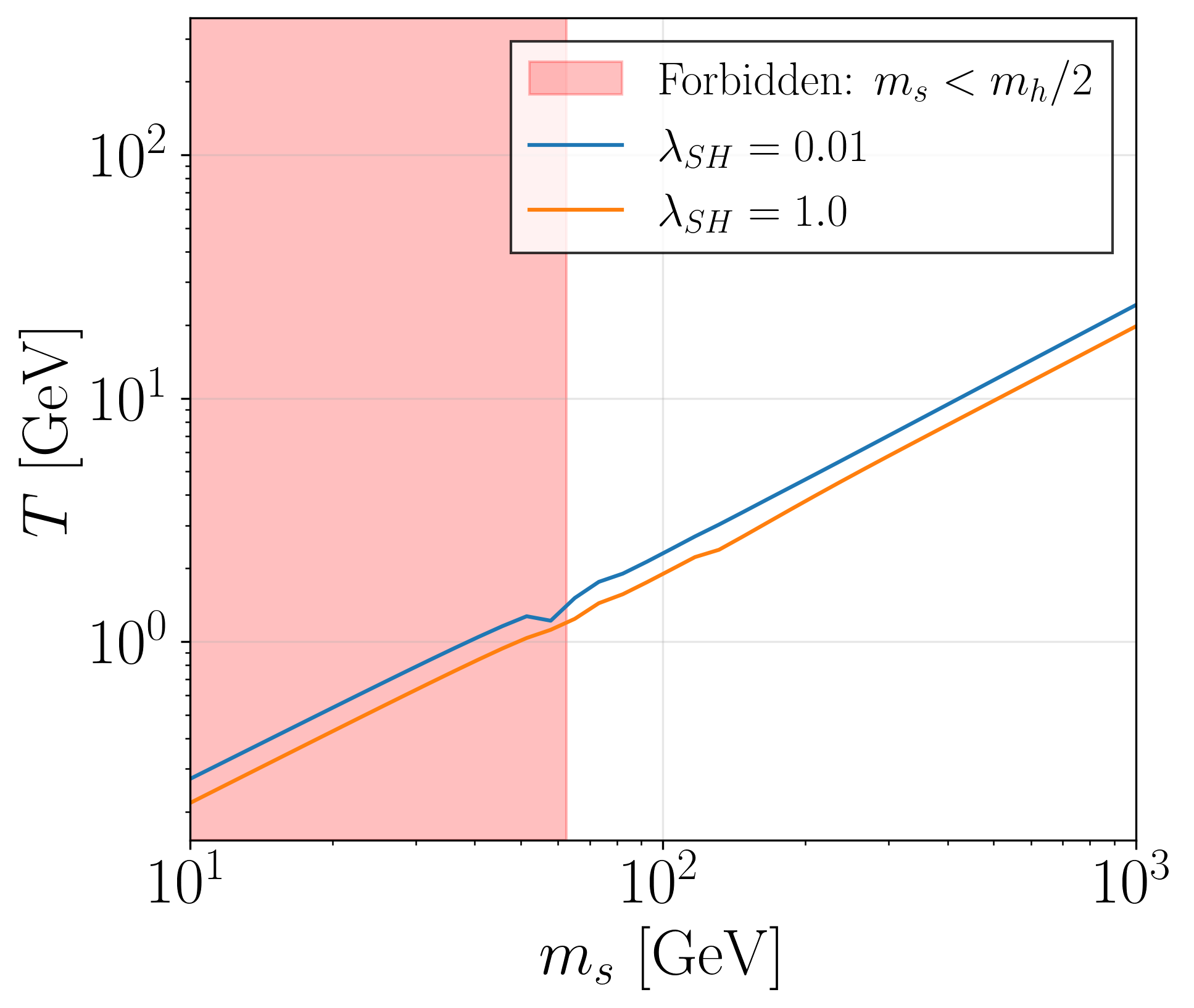}
	\caption{Minimum temperature required to thermalize the pseudoscalar mediator $s$ as a function of the Higgs-portal coupling $\lambda_{hs}$ and the pseudoscalar mass $m_s$. The red region indicates the bounds from Higgs invisible decays \cite{ATLAS:2022yvh}. These results were obtained using \texttt{micrOMEGAs}.}
	\label{plot:thermalization_s}
\end{figure}
\begin{eqnarray}
 \Gamma_s(T) = n_{s}^\text{eq}(T) \langle \sigma v \rangle_{ss\rightarrow \text{SMSM}},
\end{eqnarray}
where $n_{s}^{\rm eq}(T)$ is the equilibrium number density of pseudoscalar particles, and $\langle \sigma v \rangle_{ss\rightarrow \rm SM\,SM}$ denotes the thermally averaged cross section for the annihilation process. The thermalization condition is given by
\begin{eqnarray}\label{eq:terma2}
    2\Gamma_s(T) > 3H(T)\, .
\end{eqnarray}
For freeze-in to be correctly computed, the mediator $s$ must already be in thermal equilibrium with the SM bath by the time reheating completes. Therefore the condition $2\Gamma_s(T)>3H(T)$ must be satisfied at (or before) $T=\Trh$; otherwise $s$ would never thermalize during the epoch relevant for DM production. The temperature at which this condition first holds sets a minimum reheating temperature, and we identify this lower bound with $\Trh$ in the following analysis.

Thereby, Eq.~\ref{eq:terma2} leads to a lower bound on the reheating temperature as a function of the model parameters, as shown in Fig.~\ref{plot:thermalization_s}. For the subsequent analysis in this work, we identify this lower temperature bound with the reheating temperature $\Trh$.
\begin{figure}[t]
	\centering
\includegraphics[width=0.4\textwidth]{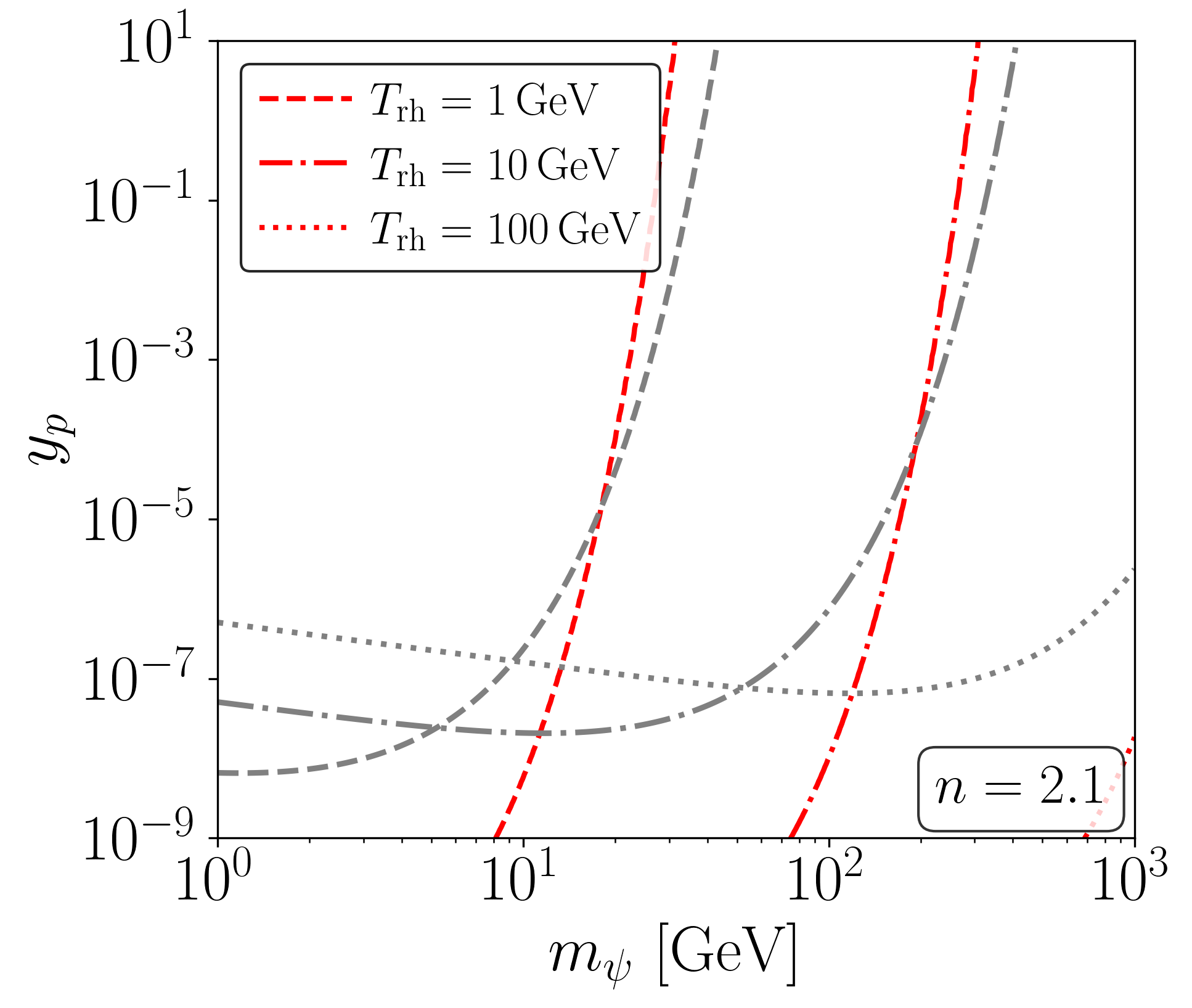}
\includegraphics[width=0.4\textwidth]{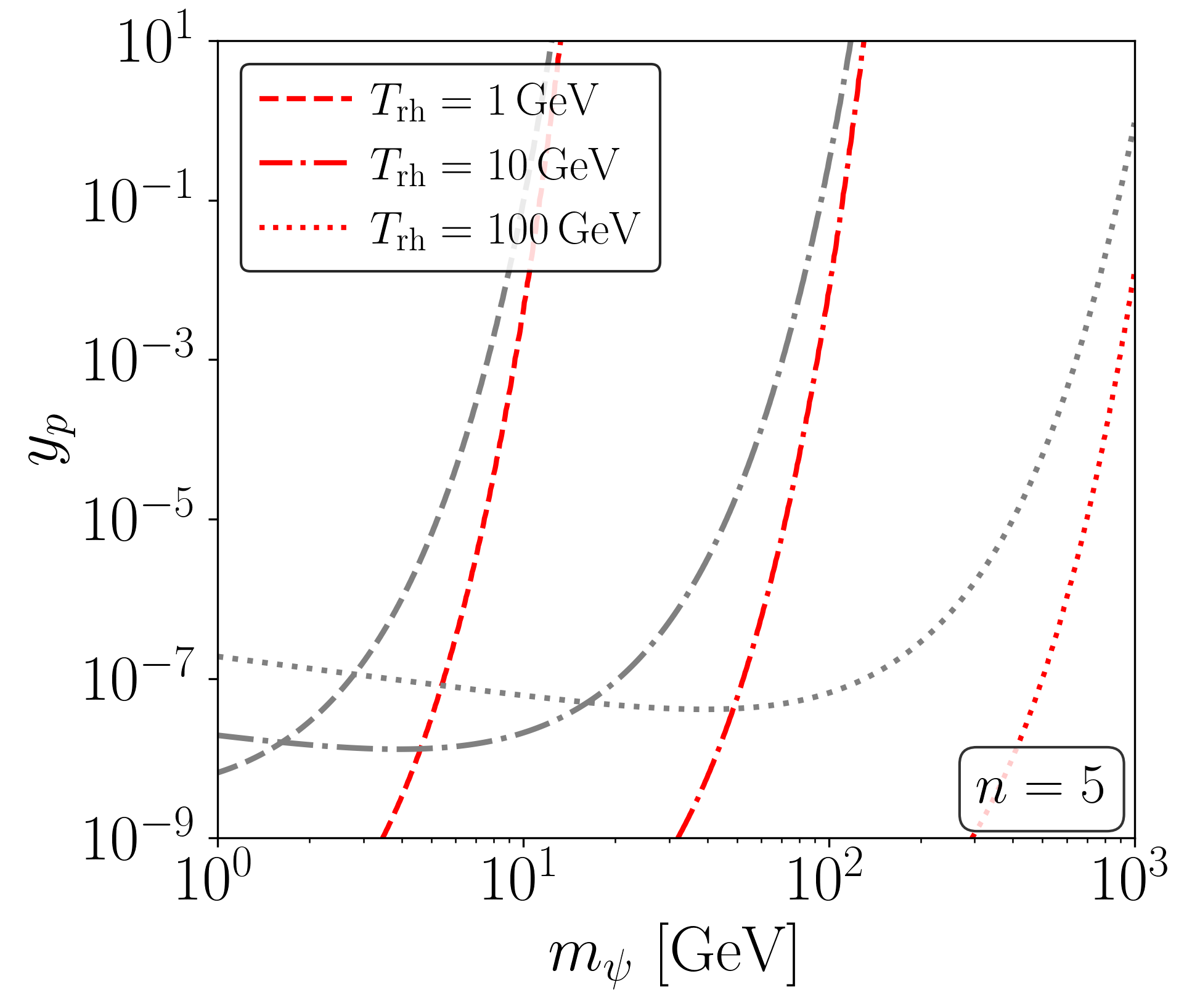}
	\caption{Thermalization bounds for different values of $n = m_s/m_\psi$ (thermalization occurs above the gray lines) and contours of the correct relic abundance via freeze-in for different reheating temperatures $\Trh$ (red lines).} 
	\label{plot:thermalization}
\end{figure}
\subsection{Fermion non-thermalization}
For a fixed reheating temperature $\Trh$, if the coupling $y_p$ is too large, the dark matter particle $\psi$ could thermalize. To ensure that $\psi$ remains out of equilibrium, we impose the condition that its interaction rate $\Gamma_\psi$ stays below the Hubble expansion rate $H$ at all relevant temperatures. The dominant process contributing to the thermalization of $\psi$ is the decay of $s$ into $\bar{\psi}\psi$. The non-thermalization condition reads
\begin{eqnarray}
 2\, n_{\psi}^\text{eq}\ev{\sigma v}_{\psi\bar\psi\rightarrow s} < 3H,
\end{eqnarray}
where $n_\psi^{\rm eq}(T)$ is the equilibrium number density of $\psi$. This condition then leads to an upper bound on the coupling $y_p$ to ensure that $\psi$ does not thermalize:
\begin{eqnarray}\label{term_1}
 y_p < \left(\frac{12\pi H(T) n_\psi^\text{eq}(T)}{m_s n_s^\text{eq}(T) \sqrt{1 - \frac{4m_\psi^2}{m_s^2}}}\right)^{1/2}.
\end{eqnarray}
This condition must be satisfied for all temperatures up to the reheating temperature $\Trh$. In Fig.~\ref{plot:thermalization}, we show the contours of the correct relic abundance (red lines) for different values of $\Trh$ and $n \equiv m_s/m_\psi$, while the gray lines indicate the boundary of thermalization for the fermion DM. That is, the region below each gray line guarantees that $\psi$ remains out of equilibrium. As expected, for larger $n$, the thermalization bounds become less stringent for the same $\Trh$, due to the stronger suppression of $n_s^{\rm eq}$ in the denominator of Eq.~\ref{term_1}.

From Fig.~\ref{plot:thermalization} we also see that achieving sizable values of $y_p$ while maintaining the non-thermalization condition favors higher $n$. As we discuss in the next section, larger $y_p$ values are generally required to probe these scenarios in direct detection experiments. We will return to the (non-)thermalization results later.

\section{Direct detection}\label{sec:dd}
\begin{figure}[t]
    \begin{subfigure}{.49\textwidth}
    \centering
        \includegraphics[width=0.8\linewidth]{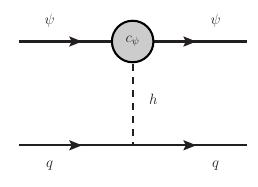}
        \caption{}
        \label{fig:dd_diagram}
    \end{subfigure}
    \begin{subfigure}{.49\textwidth}
        \centering
        \includegraphics[width=0.9\textwidth]{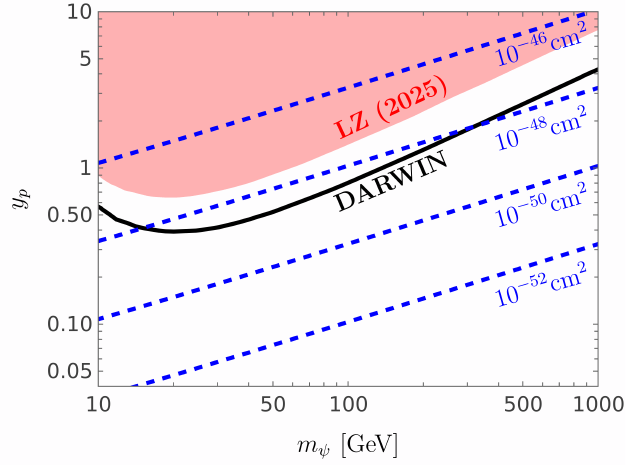}
        \caption{}
        \label{fig:SI-contours}
    \end{subfigure}
	\caption{(a) Leading contribution to the DM--Nucleon scattering cross section.
    (b) Contours of $\sigma_\text{SI}$ in the $m_\psi$--$y_p$ plane for $m_s = 2 m_\psi$ (dashed blue lines). The red region corresponds to the current exclusion limits from LZ
    \cite{LZ:2024zvo} and the black line to the projected sensitivity of DARWIN \cite{DARWIN:2016hyl}.} 
\end{figure}

At leading order the DM--nucleon interaction is mediated by t-channel scalar exchange of a Higgs boson via the diagram
in Fig.~\ref{fig:dd_diagram}.
The DM--Higgs coupling is induced purely through radiative corrections and at one-loop order reads (cf. Appendix.~\ref{app:loop})
\begin{align}
    c_\psi = \frac{y_p^2 \lambda_{hs} v}{16\pi^2 m_\psi} g\Big(\frac{m_s^2}{m_\psi^2}\Big)\, ,
\end{align}
with $g(r)$ given in Eq.~\ref{eq:DM-Higgs-LF}. The resulting (spin-independent)
direct detection cross section is given by \cite{Yaguna:2021rds}
\begin{align}
    \sigma_\text{SI} = \frac{\mu_{N\psi}^2}{\pi} \Big(\frac{m_N c_\psi}{M_h^2 v}\Big)^2 f_N^2\, ,
\end{align}
where $M_h\approx 125$ GeV denotes the Higgs mass, $m_N\approx 0.9$ GeV the nucleon mass,
$\mu_{N\psi}=m_N m_\psi / (m_N+m_\psi)$ and $f_N\approx 0.3$ \cite{Cline:2013gha}.

In Fig.~\ref{fig:SI-contours} we show contours of the SI cross section in the $m_\psi$--$y_p$ plane 
(for fixed scalar masses $m_s=2m_\psi$) as well as, for reference,
the current upper bounds obtained by LZ \cite{LZ:2024zvo} and projected limits from DARWIN \cite{DARWIN:2016hyl}. 
We note that, since $g(r)$ decreases monotonically, the SI cross section generally becomes smaller
for fixed $y_p$, if the splitting between $m_\psi$ and $m_s$ increases. In the scenarios considered here,
the direct detection constraints become relevant for couplings $y_p \gtrsim 0.5$.

\begin{figure}[t]
	\centering
\includegraphics[width=0.45\textwidth]{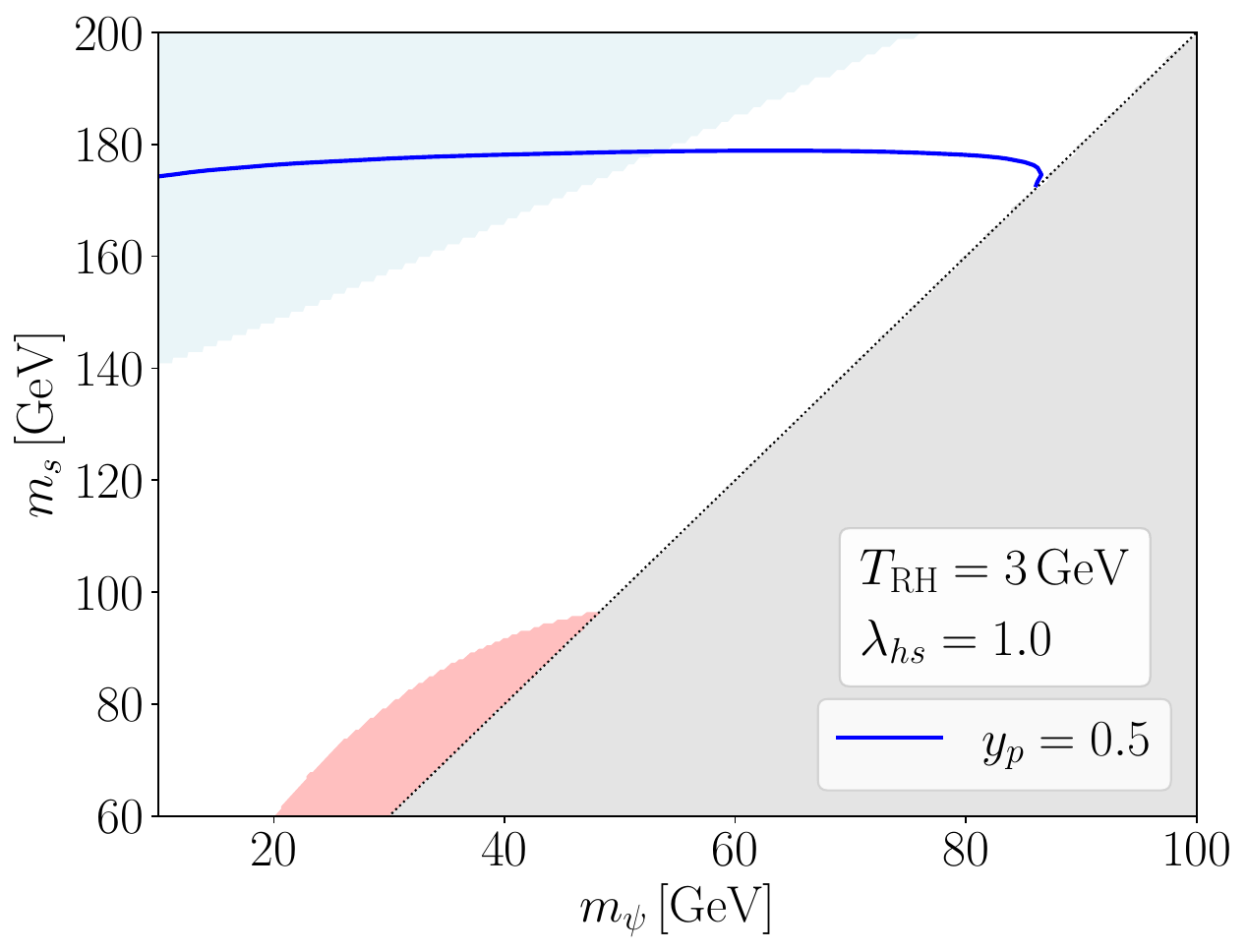}
\includegraphics[width=0.45\textwidth]{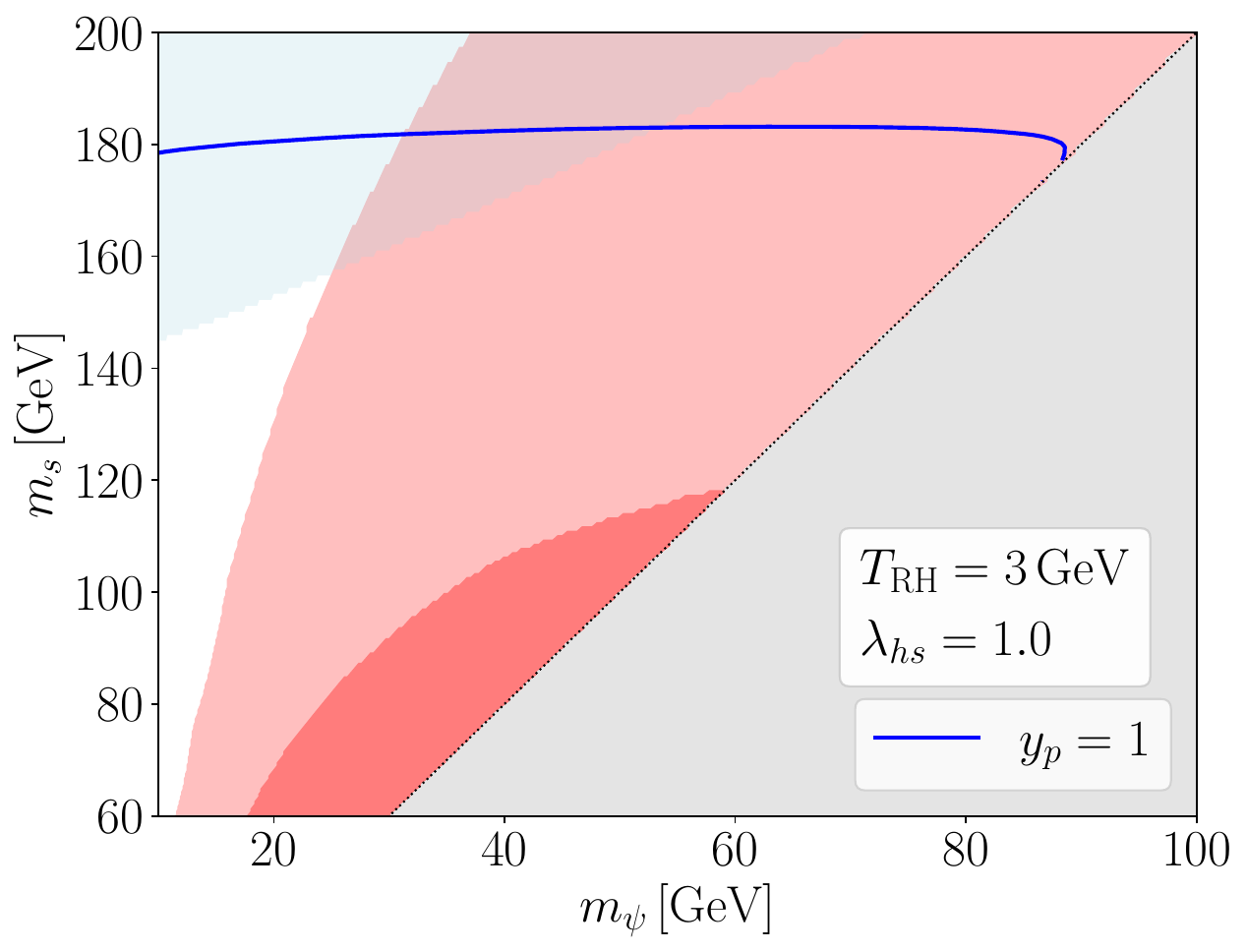}
\includegraphics[width=0.45\textwidth]{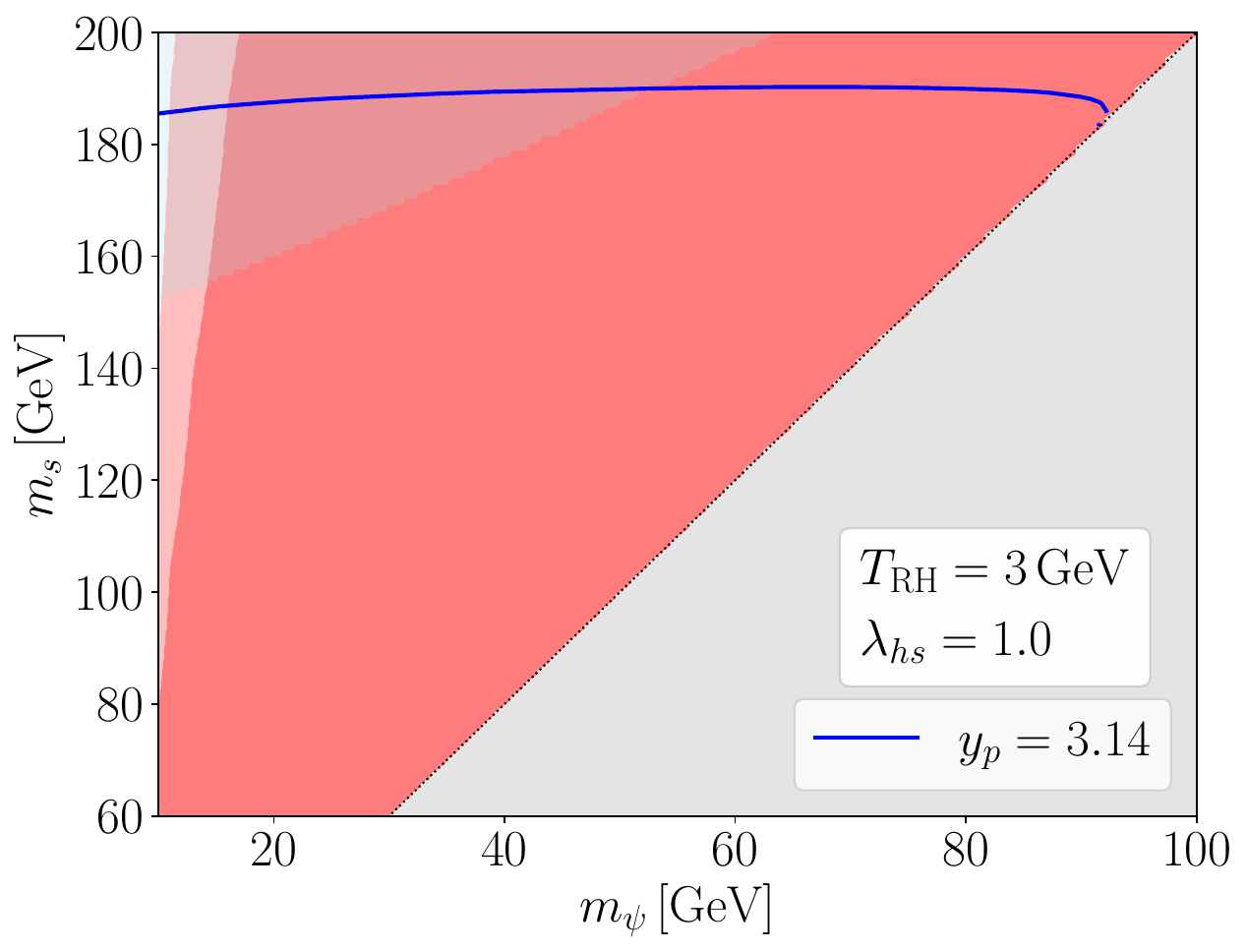}
\includegraphics[width=0.45\textwidth]{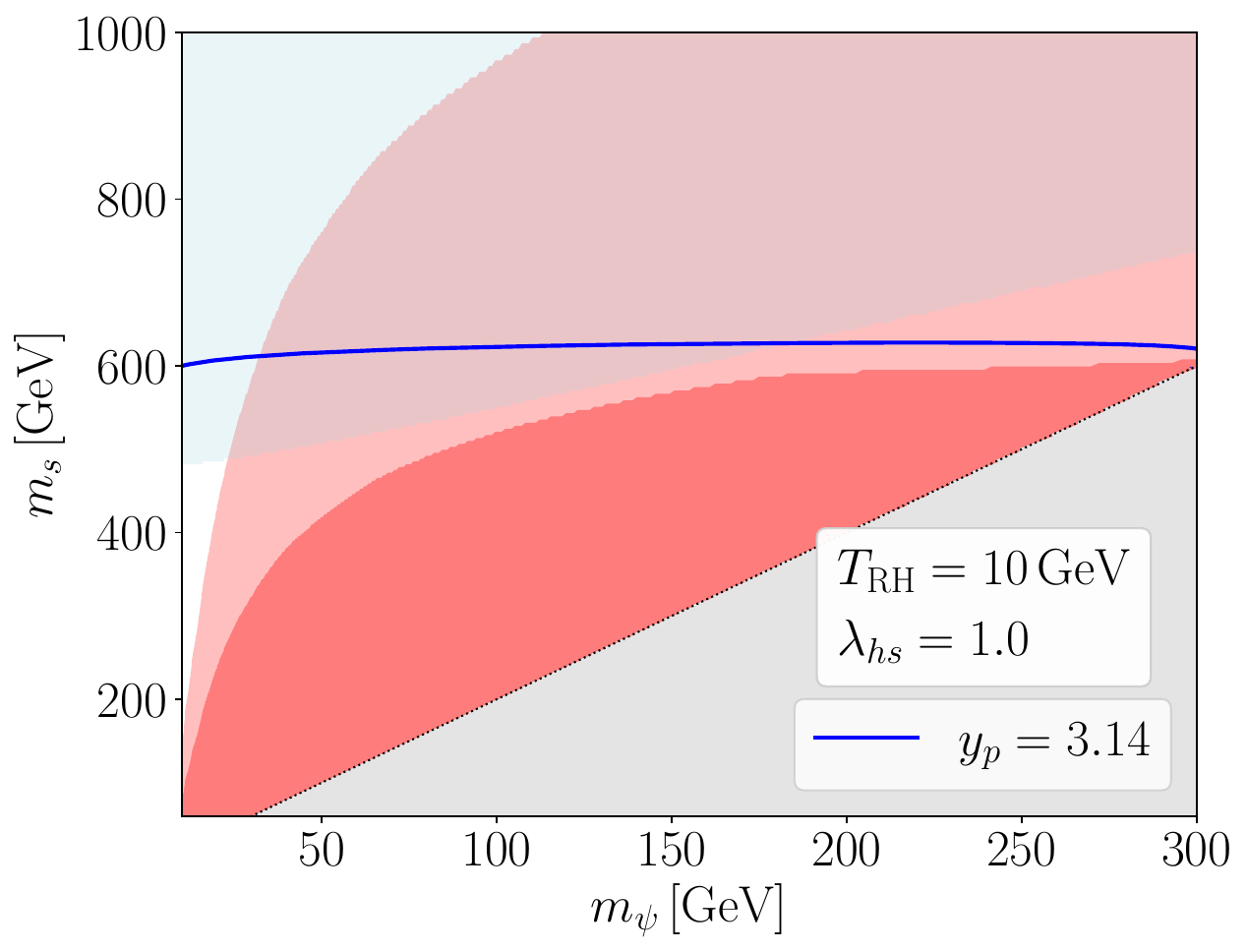}
	\caption{Region of parameter space showing the contours of correct relic abundance (blue lines), exclusion by LZ (dark red) and DARWIN projections (light red), and the region where $\psi$ remains out-of-thermal equilibrium (light blue region). The gray region is not part of our study since $m_s < 2m_\psi$.} 
	\label{fig:full}
\end{figure}

Fig.~\ref{fig:full} summarizes with a few benchmark points the interplay between the relic-density requirement, the non-thermalization condition of the fermionic dark matter particle, and present and future direct-detection bounds. The blue lines indicate the contours yielding the correct relic abundance, while the light-blue region indicates where $\psi$ never enters thermal equilibrium (see Eq.~\ref{term_1}), thereby remaining as a FIMP. The dark- and light-red regions show the current exclusion from LZ and the projected sensitivity of the future DARWIN experiment~\cite{DARWIN:2016hyl}.

For cases with $\Trh = 3~\mathrm{GeV}$, the viable FIMP DM parameter space—i.e., where the blue contours lie inside the light-blue region—is already partially tested by LZ, but only in the regime of large couplings, $g_p \sim \pi$. For lower values of $g_p$, only DARWIN will be able to probe this region; however, as shown in the top-left panel, sufficiently small couplings lie entirely beyond the reach of both experiments. In the bottom-right panel, we present another benchmark $\Trh = 10$ GeV, where even relatively large $g_p$ cannot be tested by LZ, while DARWIN retains sensitivity. Finally, we note that all the results shown satisfy the thermalization condition for the pseudo-scalar mediator $s$ (see Sec.~\ref{subsec:pseudo_therm}). 
 
\section{Discussion and Summary}\label{sec:conclusions}
We have studied the freeze-in production of fermionic dark matter via the decay of a thermal pseudo-scalar mediator in scenarios with a low reheating temperature. We derived the Boltzmann equation governing the evolution of the dark matter number density and computed the relic abundance, taking into account the conditions required for a sufficiently strong Higgs-portal coupling to thermalize the mediator~$s$, while keeping the Yukawa coupling $y_p$ small enough to preserve the FIMP paradigm, i.e.\ ensuring that the dark matter fermion $\psi$ never thermalizes.

In these two-field Higgs-portal models, dark matter does not interact with the Higgs boson at tree level. We therefore computed the finite one-loop induced effective Higgs coupling and derived the corresponding spin-independent direct detection cross section. Our results show that the model is already constrained by current direct detection searches, such as LZ, and it will be tested by future ones such as Darwin. However, regions of parameter space remain untested even under strong DARWIN projections, then suggesting other complementary ways to test these regions.

The results presented here can be straightforwardly extended to related scenarios. For instance, two-field models with scalar couplings of the form $\mathcal{L} \supset s\bar\psi\psi$ \cite{Yin:2024sle}, or setups based on new $Z_4$ symmetries that give rise to bilinears such as $\Big(
y_s\, \bar{\psi}^{\,c}\psi
+ y_s^{*}\, \bar{\psi}\psi^{c}
\Big)s$ (and their pseudoscalar counterparts, see \cite{Yaguna:2021rds, Yaguna:2023kyu, Cai:2015zza}) can also be embedded in a low-reheating-temperature freeze-in framework. 

Another interesting variation arises when the new (pseudo)scalar field acquires a vacuum expectation value. In this case, the dark matter fermion obtains a tree-level coupling to the Higgs boson, and the spin-independent direct detection cross section is produced already at tree level. Nevertheless, the qualitative picture remains the same: the relic density continues to be dominated by freeze-in production at low reheating temperatures \cite{Koivunen:2024vhr}, and present or future direct detection limits can be particularly strong, especially for light mediators \cite{Kahlhoefer:2017umn}. Moreover, in these setups—independently of the precise production mechanism—the resulting direct detection phenomenology takes the same generic form (see e.g.\ \cite{Garcia-Cely:2013nin, Garcia-Cely:2013wda, Anchordoqui:2015fra, Dutra:2015vca, Escudero:2016tzx, Arcadi:2017kky, DiazSaez:2023wli}).

\section{Acknowledgments}
BDS was partially funded by ANID–FONDECYT Iniciación grant No.~1125181.

\appendix

\section{Chemical equilibration of the pseudo-scalar mediator}
\label{app_termal}

The thermalization of the pseudo-scalar mediator $s$ with the Standard Model (SM) bath plays a crucial role in determining the production mechanism of dark matter in our setup. If $s$ reaches chemical equilibrium with the SM plasma at any point in the early Universe, its abundance follows the equilibrium distribution and the subsequent dynamics correspond to a thermal scenario (freeze-out). Conversely, if $s$ never thermalizes, its population is set through freeze-in production. Therefore, it is essential to determine whether $s$ attains chemical equilibrium.

\subsection{Thermalization of $s$}
The number density $n_s$ of $s$ evolves according to the Boltzmann equation
\begin{equation}
\dot{n}_s + 3H n_s 
= \langle\sigma v\rangle \left( \left(n_s^{\rm eq}\right)^2 - n_s^2 \right),
\label{eq:Boltzmann_s}
\end{equation}
where $n_{s,{\rm eq}}$ denotes the equilibrium number density, and $\langle\sigma v\rangle$ is the thermally averaged cross section for $ss \rightarrow XX$, with $X$ representing SM states.

To assess chemical equilibration, we expand Eq.~\eqref{eq:Boltzmann_s} close to equilibrium by writing
$
n_s = n_{s,{\rm eq}} + \delta n_s,
$
with $\delta n_s \ll n_{s,{\rm eq}}$. Linearizing the collision term, we obtain
\begin{equation}
n_{s,{\rm eq}}^2 - n_s^2 
= n_{s,{\rm eq}}^2 - \left( n_{s,{\rm eq}} + \delta n_s \right)^2
\simeq -2 n_{s,{\rm eq}}\, \delta n_s,
\end{equation}
such that Eq.~\eqref{eq:Boltzmann_s} becomes
\begin{equation}
\dot{\delta n_s}
\simeq - (2 \langle \sigma v \rangle\, n_{s,{\rm eq}} + 3H)\, \delta n_s,
\label{eq:relaxation}
\end{equation}
where the first term in the RHS of Eq.~\eqref{eq:relaxation} restores $n$ towards equilibrium, while the second term dilutes $n_s$ due to Hubble expansion. Chemical equilibration occurs when the relaxation rate exceeds the Hubble expansion rate,
\begin{eqnarray}
\Gamma_{\rm rel}(T) = 2 \langle\sigma v\rangle\, n_{s,{\rm eq}} > 3H,
\label{eq:chem_condition}
\end{eqnarray}
at some temperature $T$. If Eq.~\eqref{eq:chem_condition} is never satisfied, the pseudo-scalar never thermalizes and its abundance is set by freeze-in production.

\subsection{Thermalization of $\psi$}
To determine the condition of (non)thermalization of $\psi$, we assume that $s$ is in thermal equilibrium all the time. The number density of $\psi$ evolves according to the Boltzmann equation
\begin{equation}
\dot{n}_\psi + 3H n_\psi 
=  n_s^{\rm eq}\Gamma_{s\rightarrow\psi\bar\psi} - n_\psi^2\langle\sigma v\rangle,
\label{eq:Boltzmann_psi}
\end{equation}
where $n_{s,{\rm eq}}$ denotes the equilibrium number density, and $\langle\sigma v\rangle$ is the thermally averaged cross section for $\psi\bar\psi \rightarrow s$. Similarly to the previous case, we expand $n_\psi \approx n_{\psi}^\text{eq} + \delta n_\psi$. Retaining only linear terms in $\delta n_\psi$, and after some algebra, we find
\begin{eqnarray}
    \dot{\delta n_\psi} = n_{s}^\text{eq} \Gamma_{s\rightarrow \psi\bar\psi} - \left(n_{\psi}^\text{eq}\right)^2 \langle\sigma v\rangle -\left(2\langle\sigma v\rangle n_\psi^\text{eq} + 3H\right) \delta n_\psi .
\end{eqnarray}
The first two terms in the RHS cancel out because of detailed balance, and therefore, equivalently to the procedure followed in eq.~\eqref{eq:relaxation}, we obtain that the condition for non-thermalization of $\psi$ is  
\begin{eqnarray}
    2\frac{n_s^\text{eq}}{n_\psi^\text{eq}} \Gamma_{s\rightarrow \psi\bar\psi} < 3 H\, .
\end{eqnarray}

\section{One-loop details}\label{app:loop}
\begin{figure}[t]
	\centering
    \includegraphics[width=0.32\textwidth]{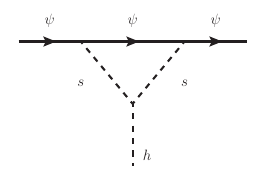}
	\caption{One-loop contribution to the effective DM--Higgs coupling.} 
	\label{DD-diagrams}
\end{figure}
Here we briefly describe the calculation of the effective DM--Higgs coupling.
Due to the absence of mixing between $s$ and $h$ this coupling is first induced at one-loop order
via the diagram in Fig.~\ref{DD-diagrams}. The corresponding amplitude is given by
\begin{align}
    \mathcal{M} &= \int \frac{d^Dk}{(2\pi)^D} \frac{ iy_p^2\lambda_{hs}v ~ [\bar{u}(q) \slashed{k} u(p)]}{
    [k^2-m_s^2][(k+p_h)^2-m_s^2][(k-p)^2-m_\psi^2]
    } \, ,
\end{align}
where $p$ and $q$ denotes the momentum of the incoming and outgoing (on-shell) DM particles
with $p^2=q^2=m_\psi^2$ and $p_h=q-p$ the momentum of the (off-shell) Higgs. After tensor reduction, the amplitude can be written in terms of Passarino-Veltman functions as
\begin{align}
\begin{split}
    \mathcal{M}&=\frac{y_p^2 \lambda_{hs} v m_\psi \bar{u}(q) u(p) }{16\pi^2[4m_\psi^2-p_h^2]} \bigg[2 B_0(p_h^2;m_s^2,m_s^2) -2B_0(m_\psi^2;m_\psi^2,m_s^2)
    +(p_h^2-2m_s^2) C_0(p_h^2,m_\psi^2,m_\psi^2;m_s^2,m_s^2,m_\psi^2)\bigg] \, , 
\end{split}
\end{align}
For direct detection, the zero momentum-transfer limit $p_h^2\to 0$ is relevant and the amplitude can be further simplified to
\begin{align}
    \mathcal{M} \overset{p_h^2=0}{=} \frac{y_p^2 \lambda_{hs} v}{16\pi^2 m_\psi} g\Big(\frac{m_s^2}{m_\psi^2}\Big) \bar{u}(q)u(p)\, ,
\end{align}
where the loop function is given by
\begin{align}\label{eq:DM-Higgs-LF}
    g(r) = \frac{r-1}{2}\ln(r) - \frac{(r-3)\sqrt{r}}{\sqrt{4-r}} \arctan\Big(\sqrt{\frac{4-r}{r}}\Big) - 1 \,.
\end{align}
From this, the spin-independent DM-nucleon cross section can be obtained using the standard results \cite{Jungman:1995df,Lin:2019uvt}.

\bibliographystyle{apsrev4-2}
\bibliography{bibliography}

\end{document}